%% file: g209.tex
\documentclass[nofootinbib,twocolumn,aps,superscriptaddress,preprintnumbers]{myRevtex4}
\pdfoutput=1

\usepackage{amsmath,amssymb,amsfonts} 
\usepackage{graphicx}
\usepackage{color}
\usepackage{xspace}
\usepackage{natbib}   
\usepackage{tabularx}

\newcommand{\figsize}{87mm}
\newcommand{\fighspace}{0.4cm}

\bibstyle{plain}
\input defs

\begin{document}

\preprint{\vbox{\hbox{BIHEP-TH-2009-004, CERN-OPEN-2009-010, LAL 09-115, arXiv:0908.4300 [hep-ph]}}}

\title{\boldmath Reevaluation of the hadronic contribution to the muon magnetic anomaly \\
        using new $\ee\to\pip\pim$ cross section data from BABAR}

\author{M.~Davier}
\affiliation{Laboratoire de l'Acc{\'e}l{\'e}rateur Lin{\'e}aire,
             IN2P3/CNRS, Universit\'e Paris-Sud 11, Orsay}
\author{A.~Hoecker}
\affiliation{CERN, CH--1211, Geneva 23, Switzerland}
\author{B.~Malaescu}
\affiliation{Laboratoire de l'Acc{\'e}l{\'e}rateur Lin{\'e}aire,
             IN2P3/CNRS, Universit\'e Paris-Sud 11, Orsay}
\author{C.Z.~Yuan}
\affiliation{Institute of High Energy Physics, Chinese Academy of Sciences, Beijing}
\author{Z.~Zhang}
\affiliation{Laboratoire de l'Acc{\'e}l{\'e}rateur Lin{\'e}aire,
             IN2P3/CNRS, Universit\'e Paris-Sud 11, Orsay}

\date{\today}

\begin{abstract}
Using recently published, high-precision \pp cross section data by the BABAR experiment
from the analysis of \ee events with high-energy photon radiation in the initial state, 
we reevaluate the lowest order hadronic contribution (\amuhadLO) to the anomalous magnetic 
moment of the muon. We employ newly developed software featuring improved data 
interpolation and averaging, more accurate error propagation and systematic validation. 
With the new data, the discrepancy between the \ee and \Tau-based results 
for the dominant two-pion mode reduces from previously $2.4\sigma$ to $1.5\sigma$ in the 
dispersion integral, though significant local discrepancies in the spectra persist. 
We obtain for the \ee-based evaluation $\amuhadLO=(695.5 \pm 4.1)\cdot10^{-10}$, where 
the error accounts for all sources. The full Standard Model 
prediction of $\amu$ differs from the experimental value by $3.2\sigma$.
\end{abstract}

\maketitle

\section{Introduction}

The Standard Model (SM) prediction of the anomalous magnetic moment of the muon, \amu, is 
limited in precision by contributions from hadronic vacuum polarisation (VP) loops. These 
contributions can be conveniently separated into a dominant lowest order (\amuhadLO) and 
higher order (\amuhadHO) parts. The lowest order term can be calculated with a 
combination of experimental cross section data involving \ee annihilation 
to hadrons, and perturbative QCD. These are used to evaluate an energy-squared 
dispersion integral, ranging from the $\piz\gamma$ threshold to 
infinity. The integration kernel strongly emphasises the low-energy part of the 
spectrum. About 73\% of the lowest order hadronic contribution 
is provided by the $\ppg$ final state.\footnote
{
   Throughout this paper, final state photon radiation is implied for the \pp final state.
} 
More importantly, 62\% of its total {\em quadratic} error stems from the $\pp$ mode, stressing 
the need for ever more precise experimental data in this channel to confirm or not the 
observed deviation of $3.8\sigma$ between SM prediction and experiment~\cite{eetaunew}.

A former lack of precision \ee-annihilation data inspired the search 
for an alternative. It was found~\cite{adh} in form of 
$\tau\to\nut+\pim\piz,\,2\pim\pip\piz,\,\pim3\piz$ spectral 
functions~\cite{aleph_old,aleph_new,cleo,opal}, transferred from the charged to 
the neutral state using isospin symmetry. During the last decade, new measurements 
of the \pp spectral function in \ee annihilation with percent accuracy became 
available~\cite{cmd2,cmd2new,snd,kloe08}, superseding or complementing older and less 
precise data. With the increasing precision, which today is on a level with the \Tau data
in that channel, 
systematic discrepancies in shape and normalisation of the spectral functions were 
observed between the two systems~\cite{dehz02,dehz03}. It was found that, when 
computing the hadronic VP contribution to the muon magnetic anomaly using the \Tau 
instead of the \ee data for the $2\pi$ and $4\pi$ channels, the observed deviation 
with the experimental value~\cite{bennett} would reduce to less than $1\sigma$~\cite{md_tau06}. 

The discrepancy with the \ee data decreased after the inclusion of new \Tau data 
from the Belle experiment~\cite{belle}, published \ee data from CMD2~\cite{cmd2new} and 
KLOE~\cite{kloe08} (superseding earlier data~\cite{kloe04}), and a reevaluation of 
isospin-breaking corrections affecting the \Tau-based evaluation~\cite{eetaunew}.\footnote
{
   The total size of the isospin-breaking correction to \amuhadLO has been estimated 
   to $(-16.1 \pm 1.9)\cdot 10^{-10}$, which is dominated by the short-distance 
   contribution of $(-12.2 \pm 0.2)\cdot 10^{-10}$~\cite{eetaunew}.
} 
In terms of \amuhadLO, the difference between the \Tau and \ee-based evaluations 
in the dominant \pp channel was found to be $11.7 \pm 3.5_{ee} \pm 3.5_{\tau+{\rm IB}}$~\cite{eetaunew} 
(if not otherwise stated, this and all following \amu numbers are given in units of $10^{-10}$), 
where KLOE exhibits the strongest discrepancy with the \Tau data (without the KLOE 
data the discrepancy reduces from $2.4\sigma$ to $1.9\sigma$). Another quantity 
for comparison, which is more sensitive to the higher-energy \pp spectrum, is 
the $\tau^-\to\pi^-\piz\nu$ branching fraction showing a difference between 
measurement and \ee prediction of $(0.64 \pm 0.10_{\tau} \pm 0.28_{ee})\%$~\cite{eetaunew}.\footnote
{
   A total isospin-breaking correction of $(+0.69 \pm 0.22)\%$ has been added to the 
   \ee prediction of the $\tau^-\to\pi^-\piz\nu$ branching fraction~\cite{eetaunew}. 
}

Recently, the BABAR Collaboration has published~\cite{babarpipi} a \ppg spectral function 
measurement based on half a million selected $\ee\to\pp\gamma(\gamma)$ events, 
where the hard photon is dominantly radiated in the initial state (ISR). It 
benefits from a large cancellation of systematic effects in the ratio $\pp\gamma(\gamma)$ 
to $\mm\gamma(\gamma)$ employed for the measurement. In this letter, we
present a reevaluation of the lowest order hadronic contribution to $\amu$ including
the new BABAR data. We deploy the new software package HVPTools~\cite{hvptools}, 
featuring a more accurate data interpolation, averaging and integration method, better 
systematic tests, and an improved statistical analysis based on the generation of 
large samples of pseudo experiments. 

\section{\boldmath$\mathsf\ee\mathsf\to\mathsf\pp$ cross section data}

The dispersion integral for the lowest order hadronic contribution reads
\beq
   \label{eq:disp}
   \amuhadLO = \frac{1}{4\pi^3}\int_{m_{\piz}^2}^\infty\!\!\!
               ds K(s)\sigma_{\ee\to{\rm hadrons}}(s)\,,
\end{equation}
where $K(s)\sim s^{-1}$~\cite{kernel}. The contribution from the light $u,d,s$ quark states 
is evaluated using exclusive experimental cross section data up to an energy of 1.8\gev, 
where resonances dominate, and perturbative QCD to predict the quark continuum beyond that
energy. In this work we only reevaluate the contributions from the $\ee\to\pp$ and 
$\pp2\piz$ channels. For all the others we refer to Refs.~\cite{dehz02,dehz03,md_tau06}.

A large number of $\ee\to\pp$ cross section measurements are available. Older measurements
stem from OLYA~\cite{barkov,E_54}, TOF~\cite{E_55}, CMD~\cite{barkov}, DM1~\cite{E_58} and 
DM2~\cite{E_59}.\footnote
{
   We do not use the data from NA7~\cite{E_56}.
} 
They are affected by an incomplete or undocumented application of radiative corrections. 
Equation~(\ref{eq:disp}) and the treatment of higher order hadronic contributions
require initial state radiation as well as leptonic and hadronic VP
contributions to be subtracted from the measured cross section data, while final state
radiation should be included. Because of lack of documentation, the latter contribution of 
approximately $0.9\%$ in the \pp channel has been added to the data, accompanied by a 
100\% systematic error~\cite{dehz02}. Initial state radiation and leptonic VP effects are 
corrected by all experiments, however hadronic VP effects are not. They are strongly 
energy dependent, and in average amount to approximately $0.6\%$. We apply this correction
accompanied by a $50\%$ systematic error~\cite{dehz02}. These FSR and hadronic VP systematic 
errors are treated as fully correlated between all measurements of one experiment, and also
among different experiments. 

More recent precision data, where all required radiative corrections have been applied
by the experiments, stem from the CMD2~\cite{cmd2new} and SND~\cite{snd} experiments at 
the VEPP-2M collider (Novosibirsk, Russia). They achieve comparable statistical errors,
and energy-dependent systematic uncertainties down to $0.8\%$ and $1.3\%$, respectively.

These measurements have been complemented by results from KLOE~\cite{kloe08} at DA$\Phi$NE 
(Frascati, Italy) running at the $\phi$ resonance centre-of-mass energy.
KLOE applied for the first time a hard-photon ISR technique to precisely 
determine the \pp cross section between $0.592$ and $0.975\gev$. The cross section data 
are obtained from a binned distribution, corrected for detector resolution and acceptance 
effects. The analysed data sample corresponds to $240\;{\rm pb}^{-1}$ integrated luminosity
providing a $0.2\%$ relative statistical error on the \pp contribution to \amuhadLO. KLOE 
does not normalise the $\pp\gamma$ cross section to $\ee\to\mm\gamma$ so that 
the ISR radiator function must be taken from Monte Carlo simulation
(\cf \cite{phokhara} and references therein). The  systematic error  assigned to this 
correction varies between $0.5\%$ and $0.9\%$ (closer to the $\phi$ peak). The total 
assigned systematic error lies between $0.8\%$ and $1.2\%$.

In a recent publication~\cite{babarpipi} the BABAR Collaboration reported measurements 
of the processes $\ee\to\ppg, \mmg$ using the ISR method at 10.6\gev centre-of-mass energy.
The detection of the hard ISR photon allows BABAR to cover a large energy range from 
threshold up to $3\gev$ for the two processes. The $\pp(\gamma)$ cross section is obtained 
from the $\pp\gamma(\gamma)$ to $\mm\gamma(\gamma)$ ratio, so that the ISR radiation function 
cancels, as well as additional ISR radiative effects. Since FSR photons are also detected, 
there is no additional uncertainty from radiative corrections at NLO level. Experimental 
systematic uncertainties are kept to 0.5\% in the $\rho$ peak region (0.6--0.9\gev), 
increasing to 1\% outside. 

\section{Combining cross section data}
\label{sec:hvptools}

The requirements for averaging and integrating cross section data are: 
($i$) properly propagate all the uncertainties in the data to the final integral 
error, ($ii$) minimise biases, \ie, reproduce the true integral as closely as 
possible in average and measure the remaining systematic error, and 
($iii$) optimise the integral error after averaging while respecting the two 
previous requirements. The first item practically requires the use of pseudo-Monte Carlo
(MC) simulation, which needs to be a faithful representation of the measurement ensemble
and to contain the full data treatment chain (interpolation, averaging, integration). 
The second item requires a flexible data interpolation method (the trapezoidal 
rule is not sufficient as shown below) and a realistic truth model used to test the 
accuracy of the integral computation with pseudo-MC experiments. Finally, the third
item requires optimal data averaging taking into account all known correlations
to minimise the spread in the integral measured from the pseudo-MC sample.

The combination and integration of the $\ee\to\pp$ cross section data is performed using 
the newly developed software package HVPTools~\cite{hvptools}.\footnote
{
   HVPTools is written in object-oriented C++ and relies on ROOT functionality~\cite{root}.
   The cross section database is provided in XML format and can be made available
   to users -- please contact the authors. The systematic errors are introduced component 
   by component as an algebraic function of mass or as a numerical value for each 
   data point (or bin). Systematic errors belonging to the same identifier (name) are
   taken to be fully correlated throughout all measurements affected. 
   So far, HVPTools has been only employed for the numerical evaluation of the most 
   important \pp (and $\pp2\piz$) parts of the dispersion integral~(\ref{eq:disp}).
} 
It transforms the bare cross section data and associated 
statistical and systematic covariance matrices into fine-grained energy bins, taking 
into account to our best knowledge the correlations within each experiment 
as well as between the experiments (such as uncertainties in radiative corrections). 
The covariance matrices are obtained by assuming common systematic error sources to 
be fully correlated. To these matrices are added statistical covariances, present for 
example in binned measurements as provided by KLOE, BABAR or the \Tau data, which 
are subject to bin-to-bin migration that has been unfolded by the experiments, thus 
introducing correlations. 

The interpolation between adjacent measurements of a given experiment uses second 
order polynomials. This is an improvement with respect to the previously 
applied trapezoidal rule, corresponding to a linear interpolation, which leads
to systematic biases in the integral (see below, and also the discussion in 
Sec.~8.2 and Fig.~12 of Ref.~\cite{dehz02}). In the case of binned data, the 
interpolation function within a bin is renormalised to keep the integral in that
bin invariant after the interpolation. This may lead to small discontinuities
in the interpolation function across bin boundaries. The final interpolation 
function per experiment within its applicable energy domain is discretised into 
small (1\mev) bins for the purpose of averaging and numerical integration. 

\begin{figure}[t]
\includegraphics[width=\figsize]{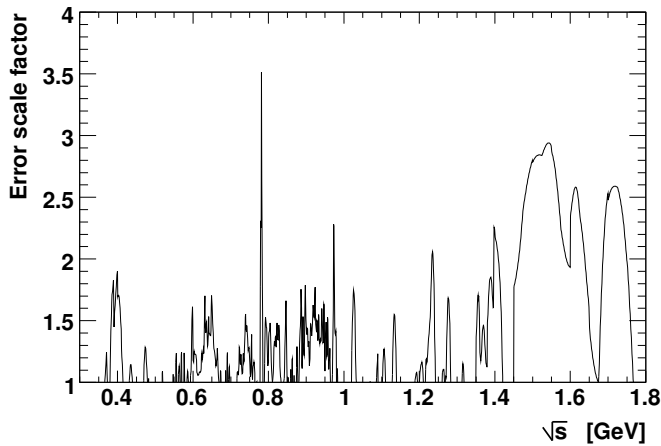}
\vspace{-0.5cm}
\caption[.]{\em Rescaling factor accounting for inconsistencies among 
            experiments versus $\sqrt{s}$ (see text). The peak around 0.4\,GeV 
            is introduced by a discrepancy between CMD2 and TOF measurements
            versus BABAR. The peaks around 0.65 and 0.74\gev are introduced by outlier
            from CMD. The sharp peak at 0.78\,GeV is due to local discrepancies 
            along the $\rho$--$\omega$ interference.
            The bump between 0.85 and 0.95\,GeV is due to a discrepancy 
            between KLOE and BABAR. Finally, between 1.45 and 1.65\,GeV
            measurements from MEA and DM2 significantly exceed the BABAR
            cross sections.
}
\label{fig:chi2}
\end{figure}

The averaging of the interpolated measurements from different experiments contributing 
to a given energy bin is the most delicate step in the analysis chain. Correlations 
between measurements and experiments must be taken into account. Moreover, the 
experiments have different measurement densities or bin widths within a given energy 
interval and one must avoid that missing information in case of a lower measurement 
density is substituted by extrapolated information from the polynomial interpolation.
To derive proper {\em averaging weights} given to each experiment, wider 
{\em averaging regions}\footnote
{
   For example, when averaging two binned measurements with unequal bin widths, a
   useful averaging region would be defined by the experiment with the larger bin width, 
   and the bins of the other experiments would be statistically merged before computing
   the averaging weights.
}  
are defined to ensure that all locally available experiments contribute to the averaging
region, and that in case of binned measurements (KLOE, BABAR, \Tau data) at least one 
full bin is contained in it. The averaging regions are used to compute weights 
for each experiment, which are applied in the bin-wise average of the original
finely binned interpolation functions.\footnote
{
   The averaging weights for each experiment are computed as follows:
   \ben
   \item pseudo-MC generation fluctuates the data points (or bins) 
         along the original measurements taking into account all known correlations; 
         the polynomial interpolation is redone for each generated pseudo MC;
   \item the averaging regions are filled for each experiment and each pseudo-MC generation
         and interpolated with second order polynomials;
   \item small (1~\mev) bins are filled for each experiment, in the energy intervals 
         covered by that experiment, using the interpolation of the averaging regions;
   \item in each small bin a correlation matrix between the experiments is computed
         from which the averaging weights are obtained.
   \een
} 
If the $\chi^2$ value of a bin-wise average\footnote
{
   The bin-wise average between experiments is computed as follows:
   \ben
   \item pseudo-MC generation fluctuates the data points (or bins) 
         along the original measurements taking into account all known correlations; 
         the polynomial interpolation is redone for each generated MC;
   \item for each generated pseudo-MC, small (1~\mev) bins are filled for each experiment, 
         in the energy intervals covered by that experiment, using the polynomial interpolation;
   \item the average and its error are computed in each small bin using the weights
         previously obtained;
   \item the covariance matrix among the experiments is computed in each small bin;
   \item $\chi^2$ rescaling corrections are computed and applied for each bin.
   \een
}
exceeds the number of degrees of freedom ($n_{\rm dof}$), the error in this averaged bin
is rescaled by $\sqrt{\chi^2/n_{\rm dof}}$ to account for inconsistencies (\cf 
Fig.~\ref{fig:chi2}). Such inconsistencies frequently occur because most experiments 
are dominated by systematic uncertainties, which are difficult to estimate. 

The consistent propagation of all errors into the evaluation of \amuhadLO is ensured 
by generating large samples of pseudo experiments, representing the 
full list of available measurements and taking into account all known correlations. 
For each generated set of pseudo measurements, the identical interpolation and 
averaging treatment leading to the computation of Eq.~(\ref{eq:disp}) 
as for real data is performed, hence resulting in a probability density distribution
for \amuhadLO (\pp), the mean and RMS of which define the $1\sigma$ allowed interval
(and which -- by construction -- has a proper pull behaviour). 
The procedure yielding the weights of the experiments can be optimised with respect 
to the resulting error on \amuhadLO. 

We have tested the fidelity of the full analysis chain (polynomial interpolation, 
averaging, integration) by using as truth representation a Gounaris-Sakurai~\cite{gounaris}
vector-meson resonance model faithfully describing the \pp data. The central values
for each of the available measurements are shifted to agree with the Breit-Wigner model,
leaving their statistical and systematic errors unchanged. The so created set of 
measurements is then analysed akin to the original data sets. The difference between 
true and estimated \amuhadLO values is a measure for the systematic uncertainty due 
to the data treatment. We find negligible bias below $0.1$ (remember 
the $10^{-10}$ unit), increasing to 0.5 (1.2 without the high-density BABAR data) when 
using the trapezoidal rule for interpolation instead of second order polynomials.

\begin{figure*}[htbp]
\begin{center}
\includegraphics[width=130mm]{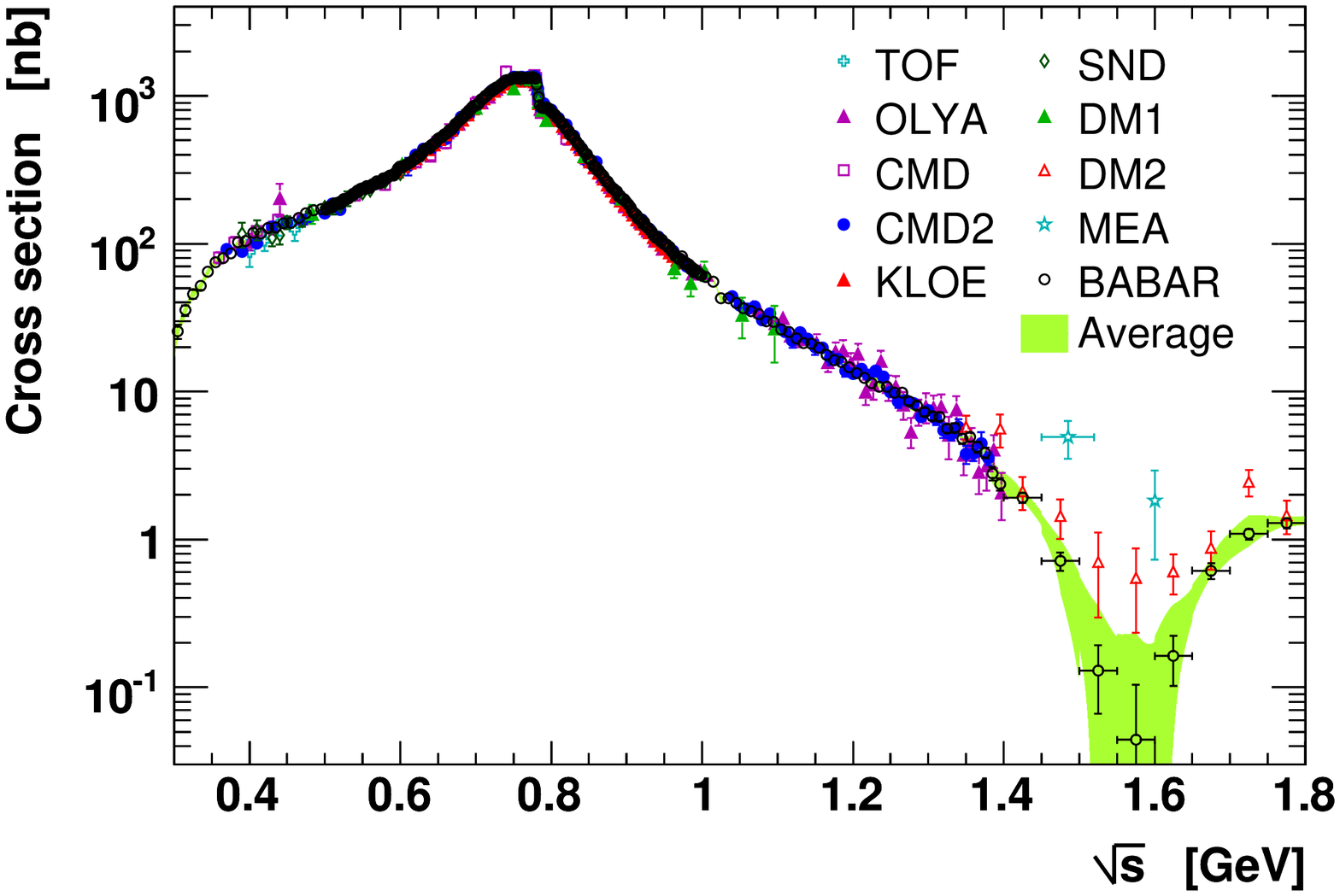}
\vspace{0.5cm}

\includegraphics[width=\figsize]{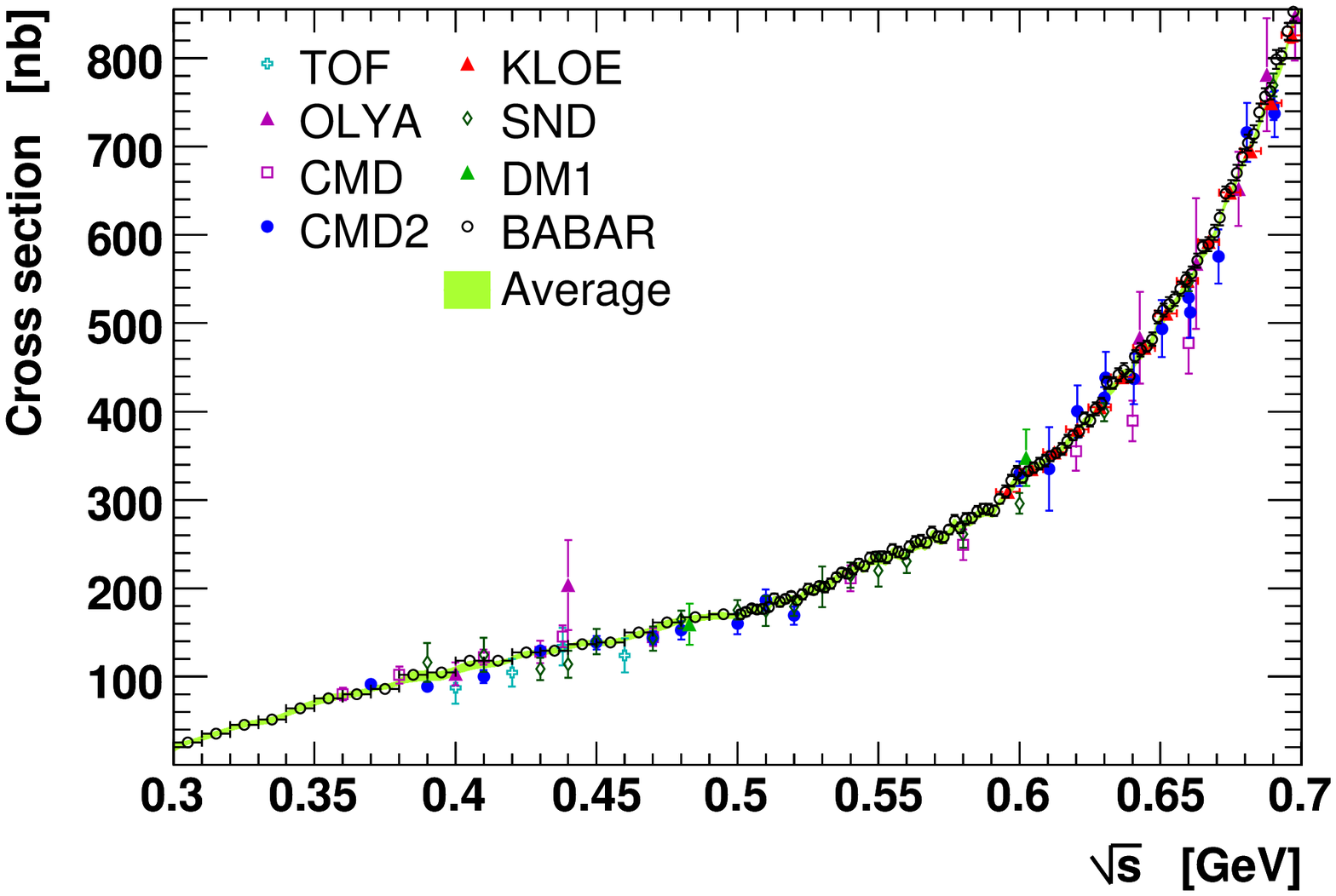}\hspace{\fighspace}
\includegraphics[width=\figsize]{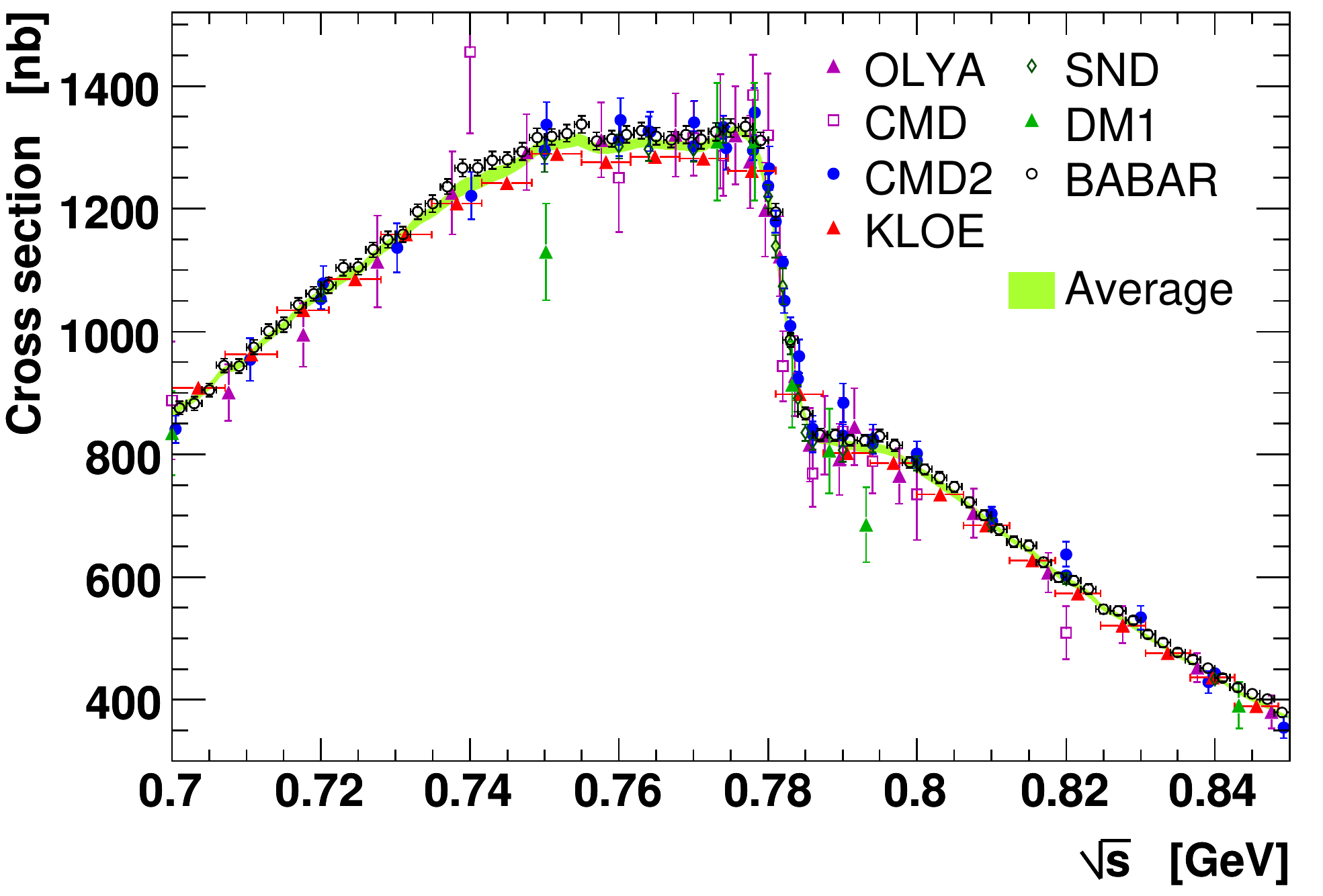}
\vspace{0.1cm}

\includegraphics[width=\figsize]{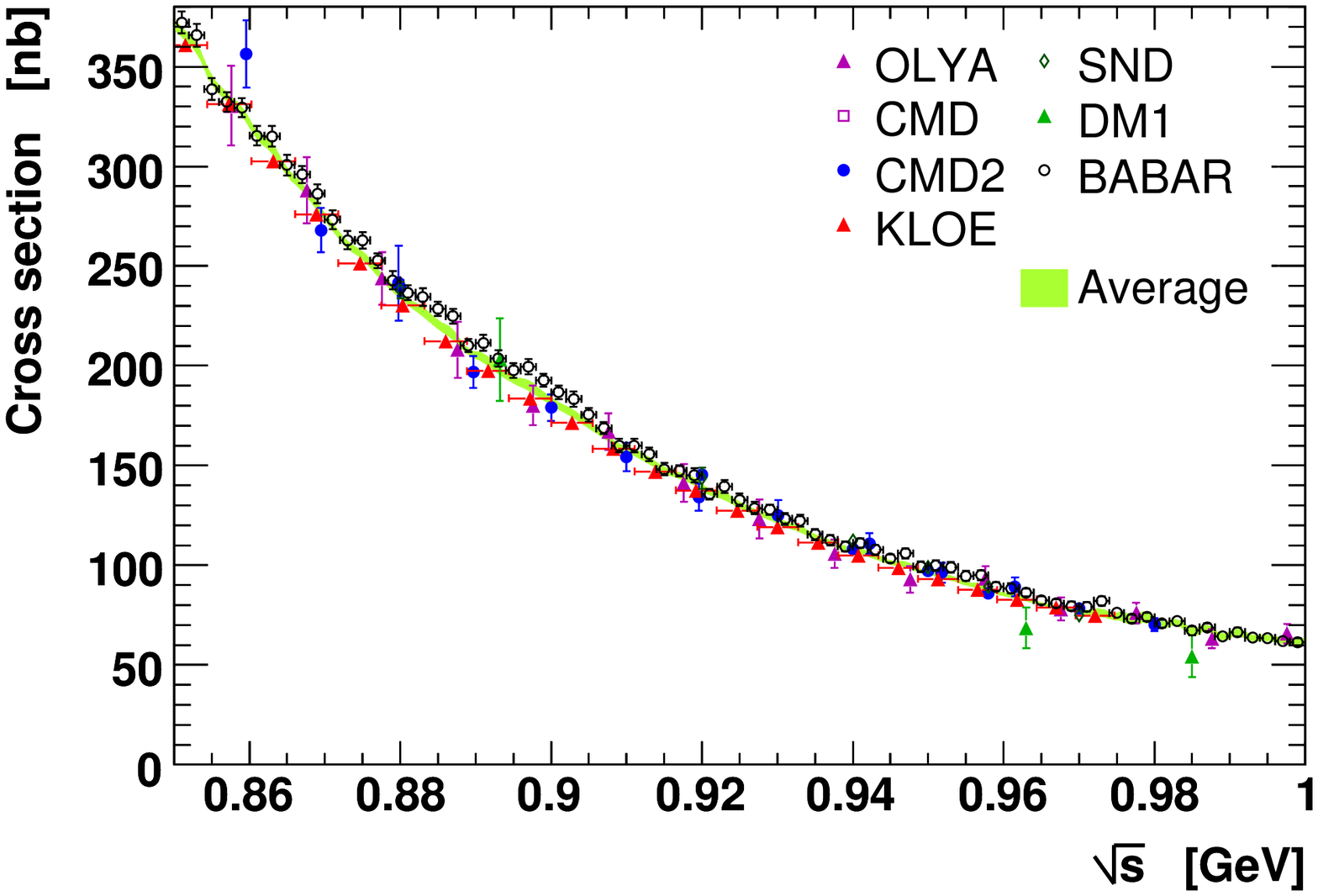}\hspace{\fighspace}
\includegraphics[width=\figsize]{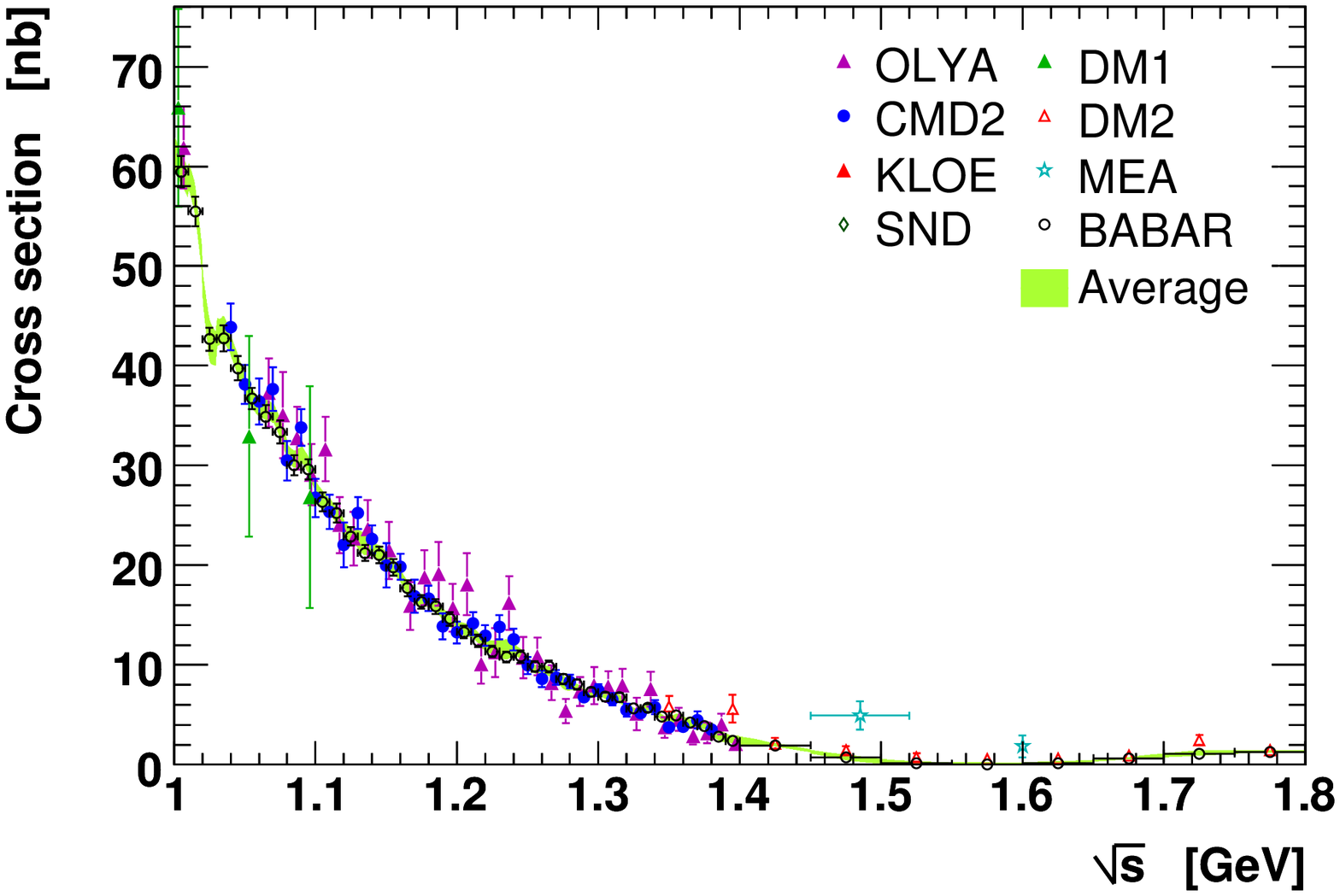}
\end{center}
\vspace{-0.3cm}
\caption[.]{\em Cross section for $\ee\to\pp$ annihilation measured by the different experiments
            for the entire energy range (top), and zoomed energy intervals (all other plots). 
            The errors bars contain both statistical and systematic errors, added in 
            quadrature. The shaded (green) band represents the average of all the 
            measurements obtained by HVPTools, which is used for the numerical integration 
            following the procedure discussed in Sec.~\ref{sec:hvptools}.}
\label{fig:xsec}
\end{figure*}

\begin{figure*}[htbp]
\begin{center}
\includegraphics[width=\figsize]{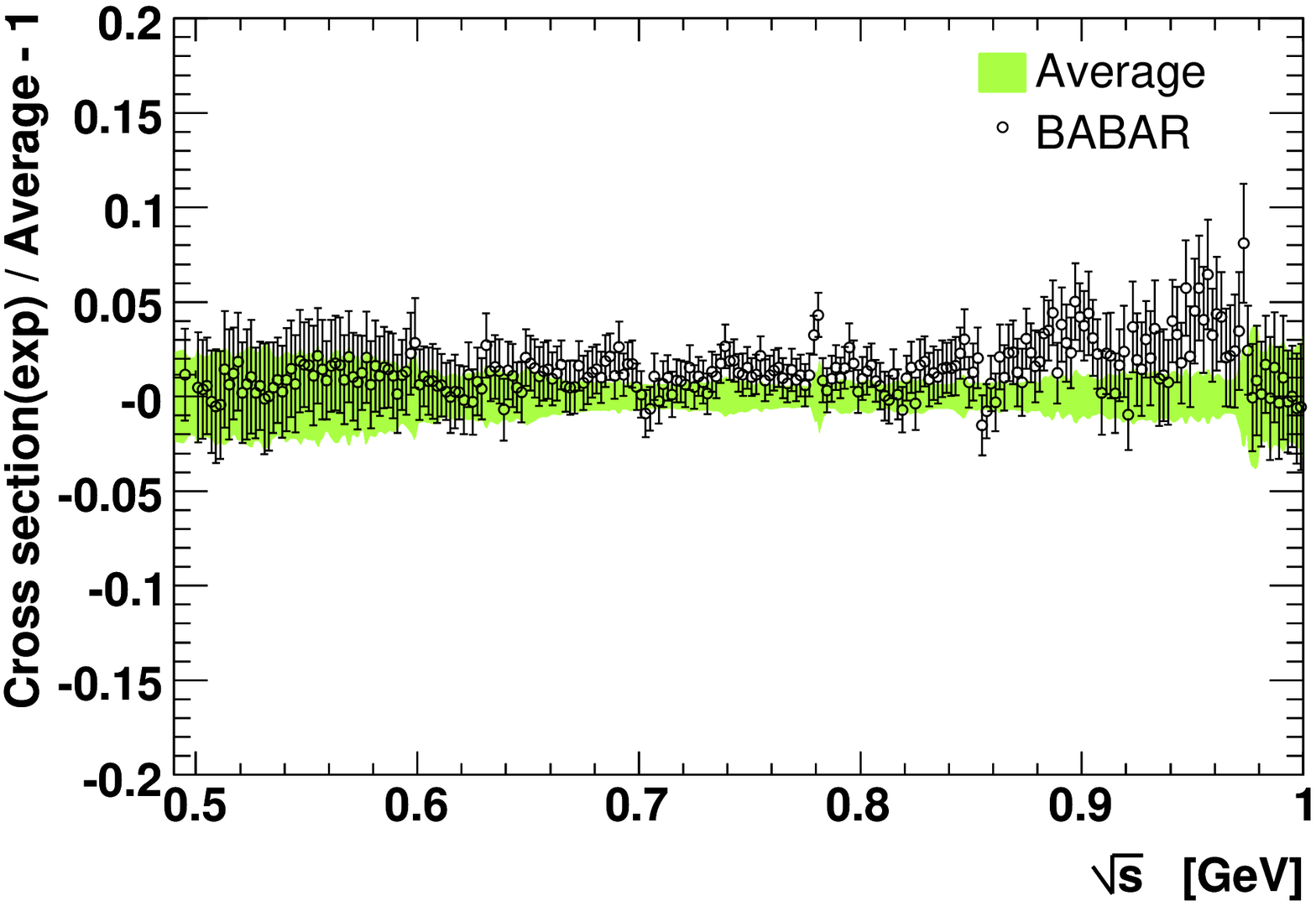}\hspace{\fighspace}
\includegraphics[width=\figsize]{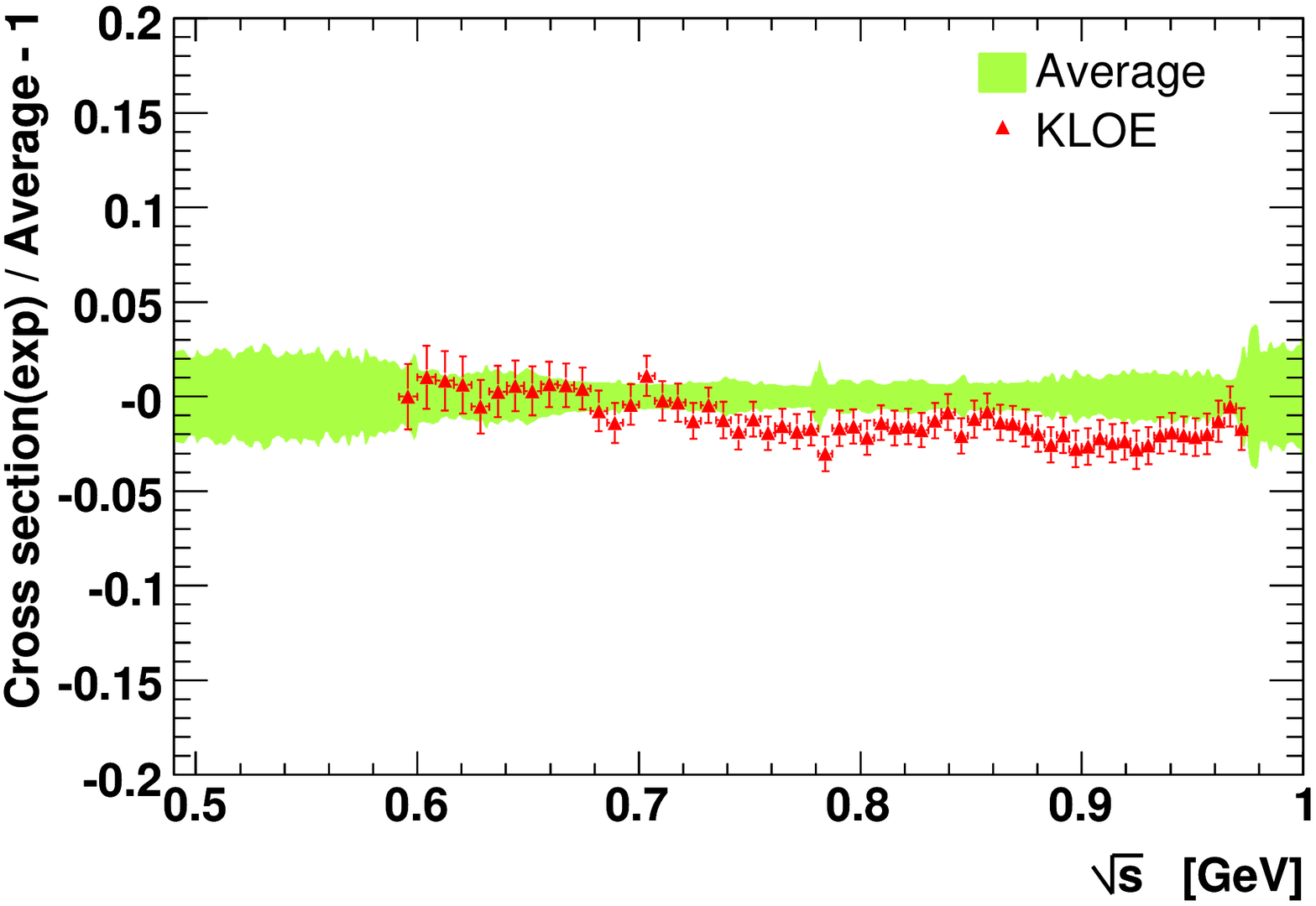}
\vspace{0.1cm}

\includegraphics[width=\figsize]{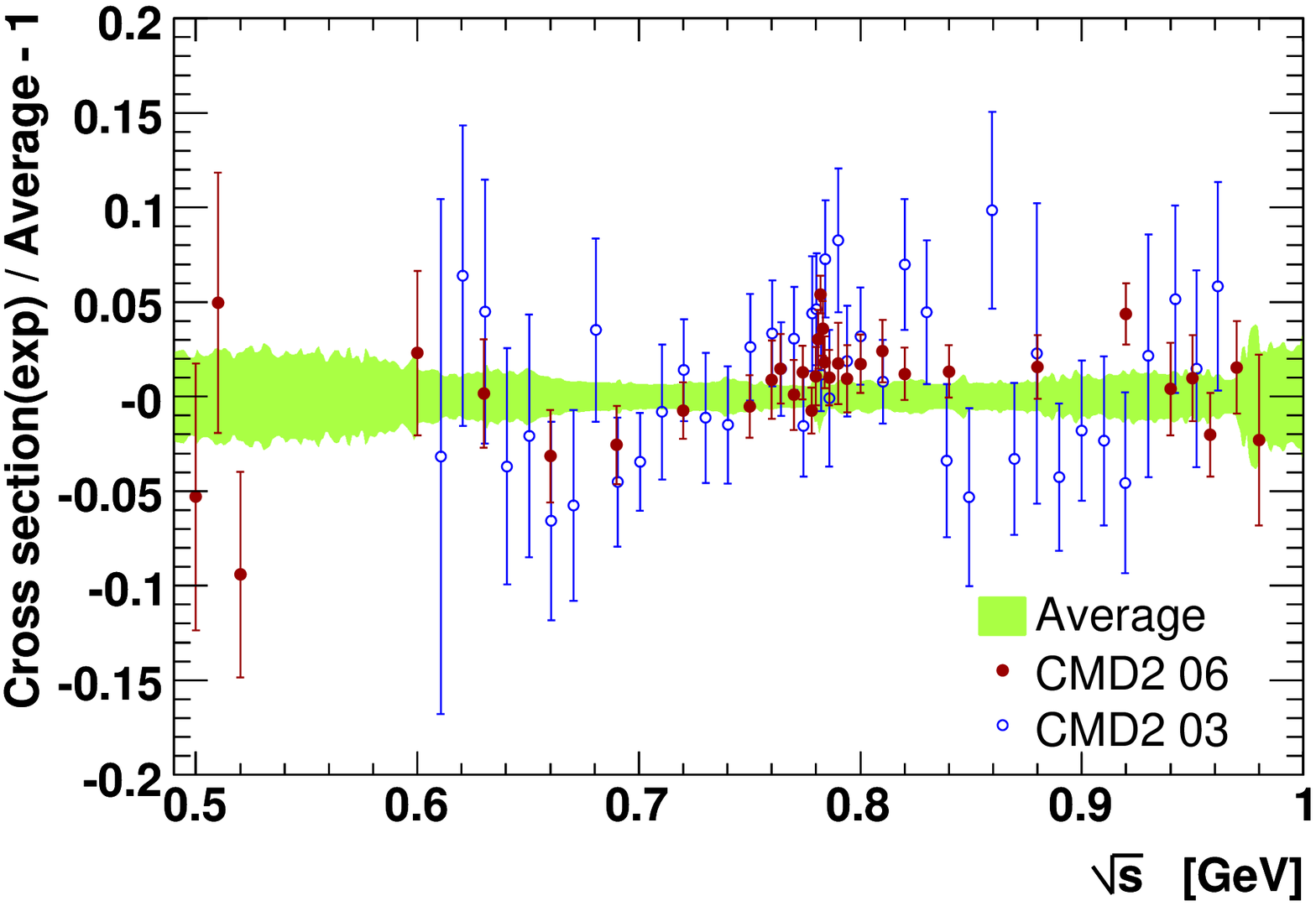}\hspace{\fighspace}
\includegraphics[width=\figsize]{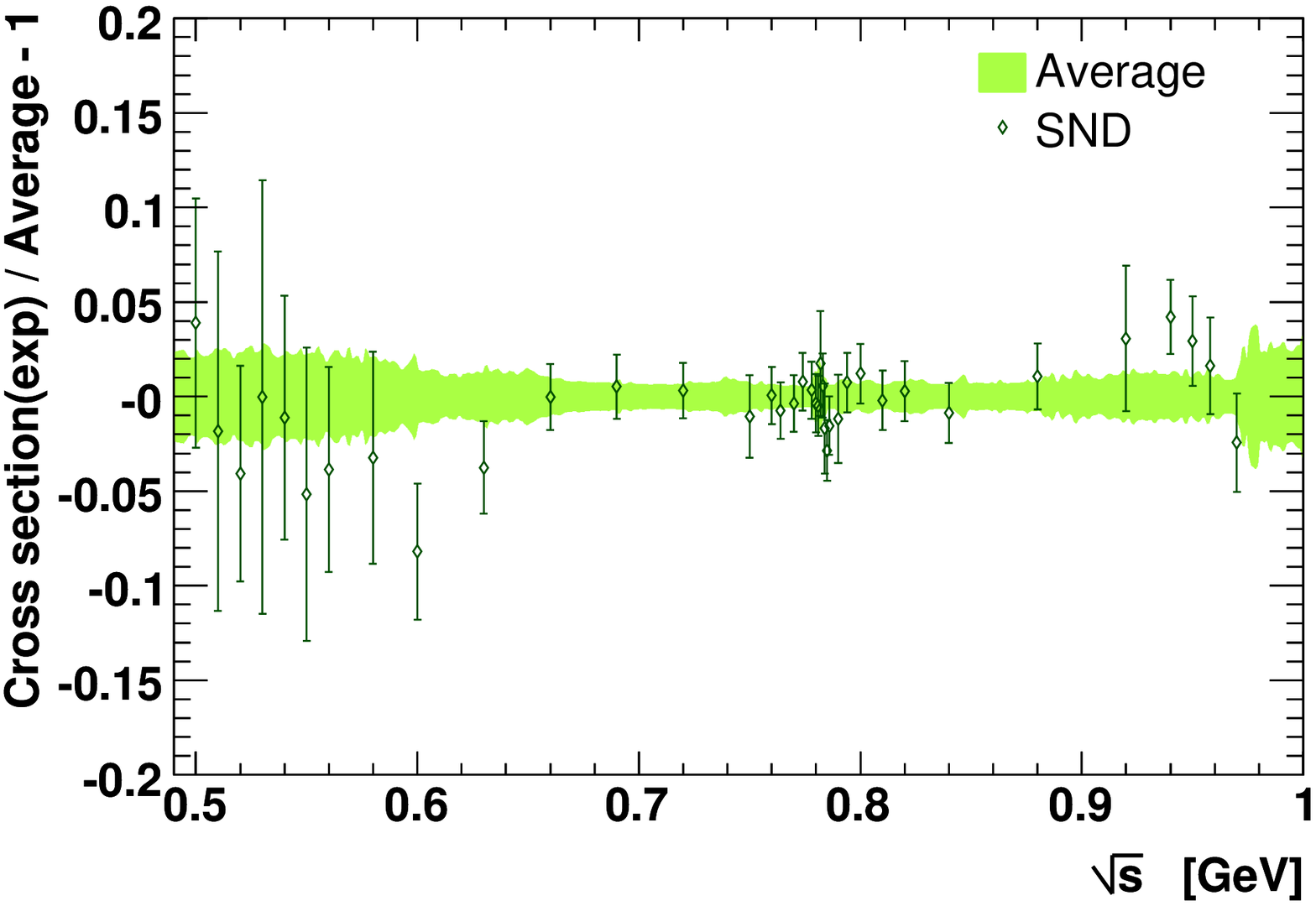}
\end{center}
\vspace{-0.3cm}
\caption[.]{\em Relative cross section comparison between individual experiments 
            (symbols) and the HVPTools average (shaded band) computed from all 
            measurements considered. Shown are BABAR (top left), KLOE (top right), 
            CMD2 (bottom left) and SND (bottom right).}
\label{fig:comp}
\vspace{1.1cm}

\includegraphics[width=\figsize]{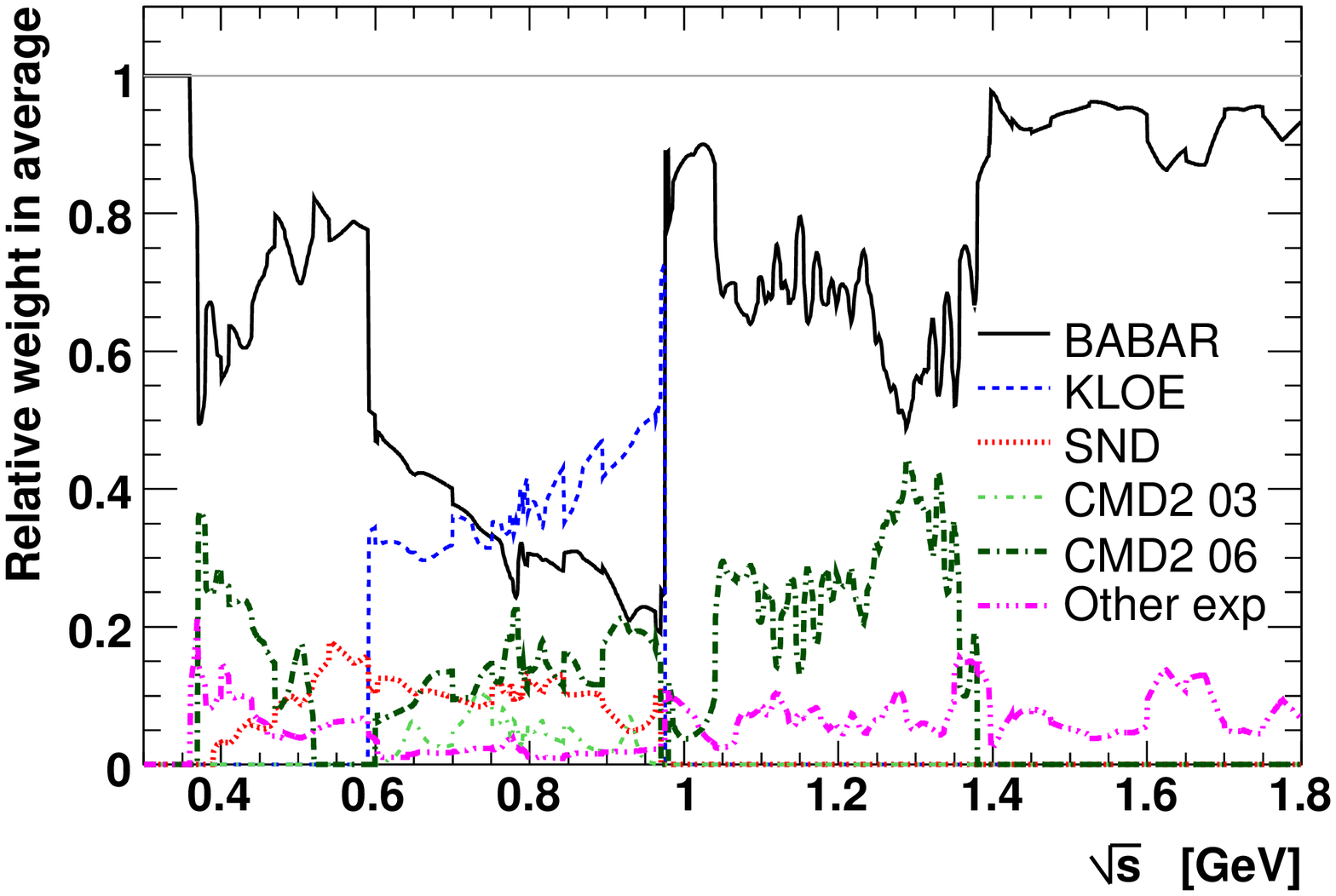}\hspace{\fighspace}
\includegraphics[width=\figsize]{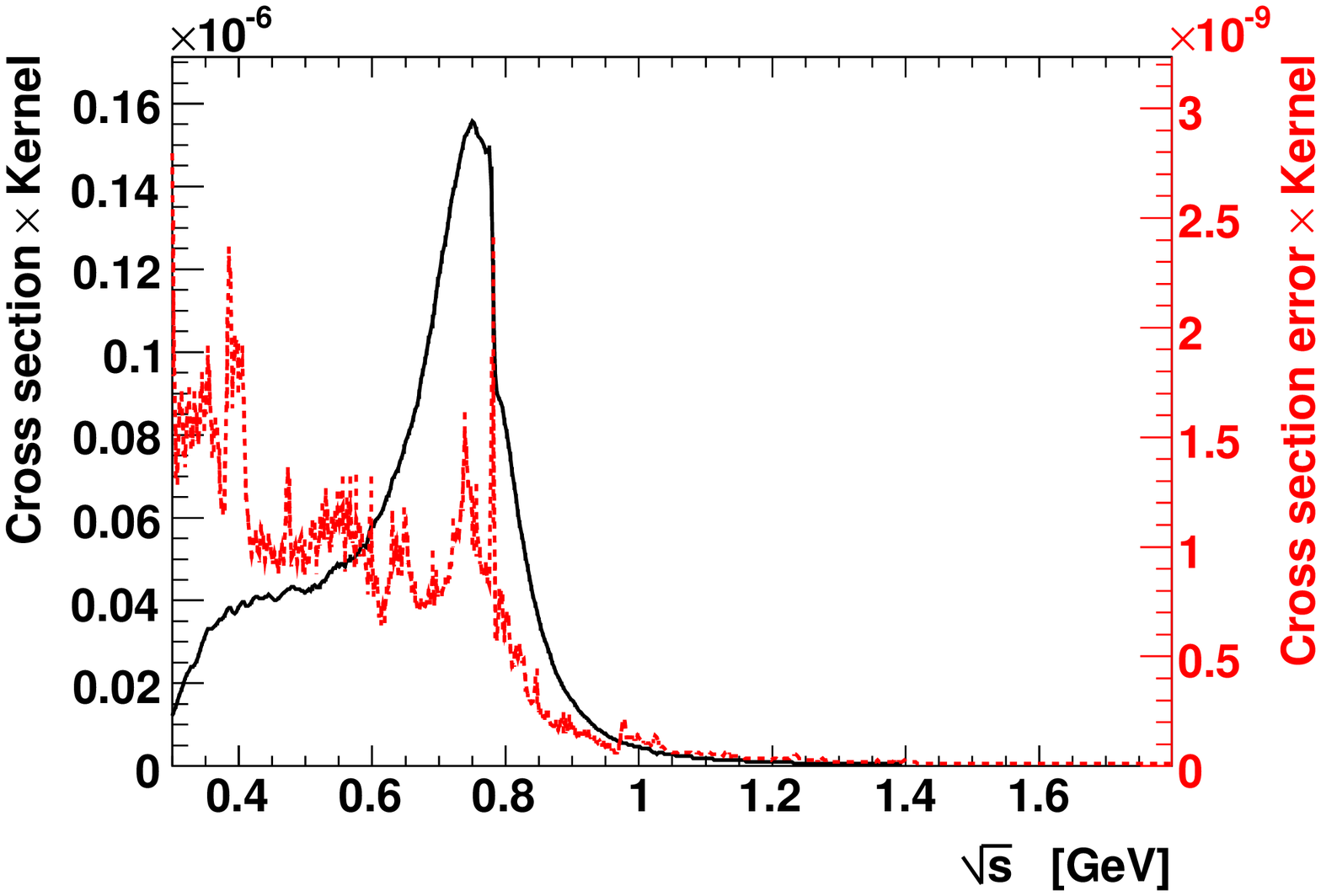}
\vspace{-0.3cm}
\caption[.]{\em \underline{Left:} relative averaging weights per experiment versus $\sqrt{s}$. 
            \underline{Right:} contribution to the dispersion integral~(\ref{eq:disp}) for the 
            combined \ee data obtained by multiplying the \pp cross section by 
            the kernel function $K(s)$ (solid line). The dashed (red) curve belonging to the right 
            axis shows the corresponding error contribution, where statistical and 
            systematic errors have been added in quadrature. Note that the information 
            conveyed by this curve is incomplete because only diagonal errors
            are shown, disregarding correlations between the cross section measurements
            which have significant influence on the integral error.}
\label{fig:weights}
\end{figure*}

The individual $\ee\to\pp$ cross section measurements (dots) and their average (shaded/green 
band) are plotted in Fig.~\ref{fig:xsec}. The error bars contain statistical and 
systematic errors. For better comparison we also plot in Fig.~\ref{fig:comp} the relative 
differences between BABAR, KLOE, CMD2, SND, and the average. Fair agreement is observed, 
though with a tendency to larger (smaller) cross sections above $\sim$$0.8\gev$ for 
BABAR (KLOE). These inconsistencies (among others) lead to the error rescaling shown 
versus $\sqrt{s}$ in Fig.~\ref{fig:chi2}. 

The left hand plot of Fig.~\ref{fig:weights} shows the weights versus $\sqrt{s}$ the different 
experiments carry in the average. BABAR and KLOE dominate over the entire energy range. 
Owing to the sharp radiator function, the available statistics for KLOE increases towards the $\phi$ 
mass, hence outperforming BABAR above $\sim$$0.8\gev$. For example, at $0.9\gev$ KLOE data 
have  statistical errors of $0.5\%$, which is twice smaller than for BABAR (renormalising
BABAR to the 2.75 times larger KLOE bins at that energy). Conversely, at $0.6\gev$ the 
comparison reads $1.2\%$ (KLOE) versus $0.5\%$ (BABAR, again given in KLOE bins which 
are about 4.2 times larger than BABAR at that energy). The experiments labelled ``other exp'' 
in the figure correspond to older data with incomplete radiative corrections. Their 
weights are small throughout the entire energy domain.

Figure~\ref{fig:weights} (right) shows versus $\sqrt{s}$ the combined $\ee\to\pp$ cross section 
multiplied by the kernel function $K(s)$ occurring in the dispersion integral~(\ref{eq:disp}). 
The kernel strongly emphasises the low-energy spectrum. The dashed (red) curve belonging 
to the right axis in the plot gives the corresponding error contribution (diagonal errors only, 
statistical and systematic errors have been added in quadrature). The peaks are introduced 
by the error rescaling and indicate inconsistencies between the measurements. The 
uncertainty in the integral is dominated by the measurements below 0.8\gev.

\section{Results}

\begin{table*}[t]
  \caption[.]{\label{tab:results}
    Evaluated $\amuhadLOpp$ contributions from the \ee data 
    for different energy intervals and experiments. Where two errors are given, the 
    first is statistical and the second systematic. Also given 
    is the \Tau-based result from Ref.~\cite{eetaunew} combining all available 
    \Tau data. The combined error has been rescaled
    to account for the inconsistency between the two evaluations.
}
\setlength{\tabcolsep}{0.0pc}
\begin{tabularx}{\textwidth}{@{\extracolsep{\fill}}lll} 
\hline\noalign{\smallskip}
  Energy range (GeV)  &  Experiment   &  $\amuhadLO[\pi\pi]$ ($10^{-10}$) \\
\noalign{\smallskip}\hline\noalign{\smallskip}
$2m_{\pi^\pm}-0.3$ & Combined \ee (fit)  & $0.55\pm 0.01$ \\ 
\noalign{\smallskip}\hline\noalign{\smallskip}
$0.30-0.63$   & Combined \ee  & $132.6 \pm 0.8 \pm 1.0$ ($1.3_{\rm tot}$) \\
\noalign{\smallskip}\hline\noalign{\smallskip}
$0.63-0.958$ & CMD2 03        & $361.8 \pm 2.4 \pm 2.1$ ($3.2_{\rm tot}$) \\
             & CMD2 06        & $360.2 \pm 1.8 \pm 2.8$ ($3.3_{\rm tot}$) \\
             & SND  06        & $360.7 \pm 1.4 \pm 4.7$ ($4.9_{\rm tot}$) \\
             & KLOE 08        & $356.8 \pm 0.4 \pm 3.1$ ($3.1_{\rm tot}$) \\
             & BABAR 09       & $365.2 \pm 1.9 \pm 1.9$ ($2.7_{\rm tot}$) \\
             & Combined \ee   & $360.8 \pm 0.9 \pm 1.8$ ($2.0_{\rm tot}$) \\
\noalign{\smallskip}\hline\noalign{\smallskip}
$0.958-1.8$  & Combined \ee   &  $14.4 \pm 0.1 \pm 0.1$ ($0.2_{\rm tot}$) \\ 
\noalign{\smallskip}\hline\noalign{\smallskip}
Total        & Combined \ee   & $508.4 \pm 1.3 \pm 2.6$ $(2.9_{\rm tot})$\\
Total        & Combined \Tau~\cite{eetaunew}
                              & $515.2\pm 2.0_{\rm exp}\pm 2.2_\BR\pm 1.9_{\rm IB}$ $(3.5_{\rm tot})$ \\
\noalign{\smallskip}\hline
\end{tabularx}
\end{table*}

A compilation of results for \amuhadLOpp for the various sets of experiments and energy 
regions is given in Table~\ref{tab:results}. 
The comparison with our previous result~\cite{eetaunew}, $\amuhadLOpp=503.5\pm3.5_{\rm tot}$, 
shows that the inclusion of the new BABAR data significantly increases the central value 
of the integral, without however providing a large error reduction. This
is due to the incompatibility between mainly BABAR and KLOE, causing an increase of
the combined error. In the energy interval between 0.63 and 0.958\gev, the discrepancy 
between the \amuhadLOpp evaluations from KLOE and BABAR amounts to $2.0\sigma$.
BABAR is the only experiment covering the entire energy region between $2m_\pi$ and
1.8\gev. Using only the BABAR data to evaluate \amuhadLOpp one finds~\cite{babarpipi} 
$514.1 \pm 2.2_{\rm stat} \pm 3.1_{\rm syst}$.

\begin{figure}[htbp]
\includegraphics[width=\figsize]{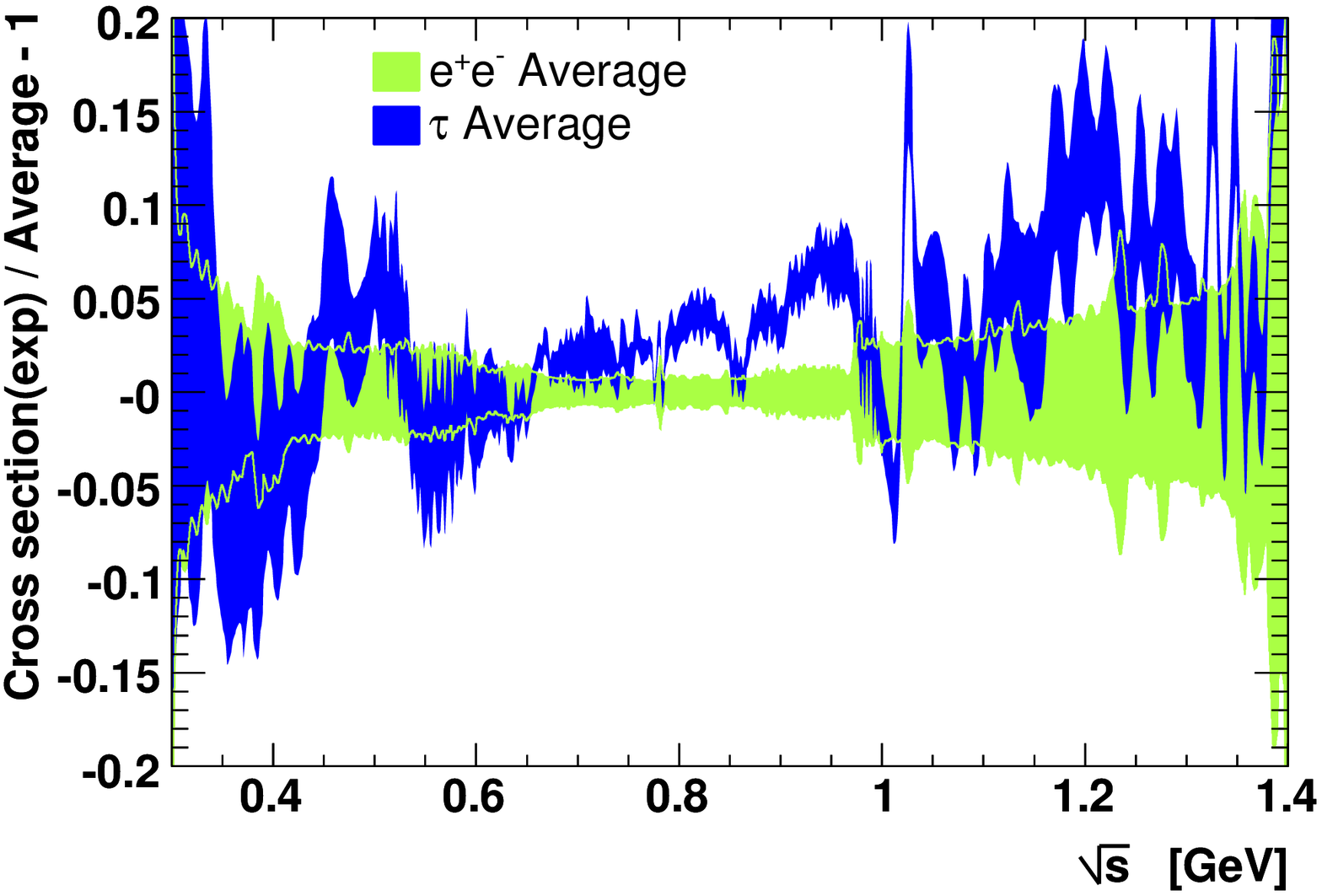}
\vspace{-0.3cm}
\caption[.]{\em Relative comparison between the combined \Tau (dark shaded) and 
            \ee spectral functions (light shaded), normalised to the \ee result.
            The apparently oscillating structure around 0.5\,GeV is due to 
            two Belle measurements fluctuating to large cross section values. 
            Clearly visible is the interference due to $\rho$--$\phi$ mixing around
            1\,GeV, which is not included in the isospin-breaking corrections applied
            to the \Tau data. It is also visible in the upper, and lower right hand 
            plots of Fig.~\ref{fig:xsec}. The deviation between 0.8 and 0.95\,GeV
            is due to the discrepancy between \Tau and KLOE data, which dominate 
            in this region (cf. Fig.~\ref{fig:weights} left).
	    Comparing the \Tau data with the combined \ee data instead of a fit to a single 
	    experiment CMD-2 limited to 1 GeV as it was done for Fig. 4 in Ref.~\cite{jegerproc} and Fig. 28 in Ref.~\cite{jeger}, we observe 
	    a reduced discrepancy, in particular between 
	    1.0GeV and 1.4GeV. We therefore disagree with the conclusion reached in these 
	    references, where the difference goes up to a factor 4, and is even in the opposite 
	    direction with respect to the one we observe.}
\label{fig:eetau}
\end{figure}
\begin{figure}[t]
\includegraphics[width=\figsize]{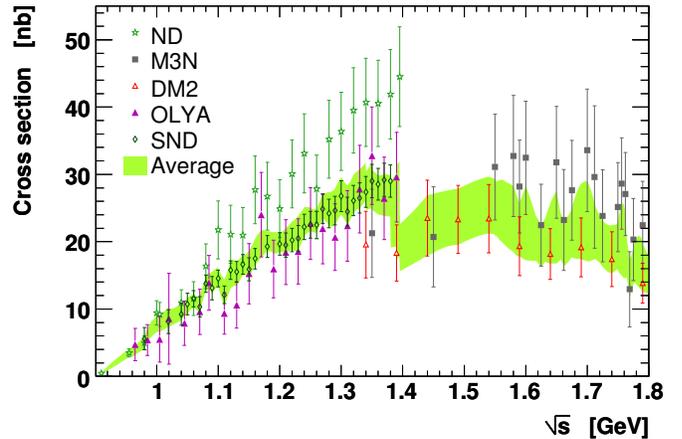}
\vspace{-0.3cm}
\caption[.]{\em Cross section measurements $\ee\to\pp2\piz$  used
            in the calculation of \amuhadLOppzz. The shaded band depicts the 
            HVPTools interpolated average within $1\sigma$ errors.
            The individual measurements are referenced in~\cite{dehz02}. }
\label{fig:4pi}
\end{figure}

\begin{figure}[htbp]
\includegraphics[width=\columnwidth]{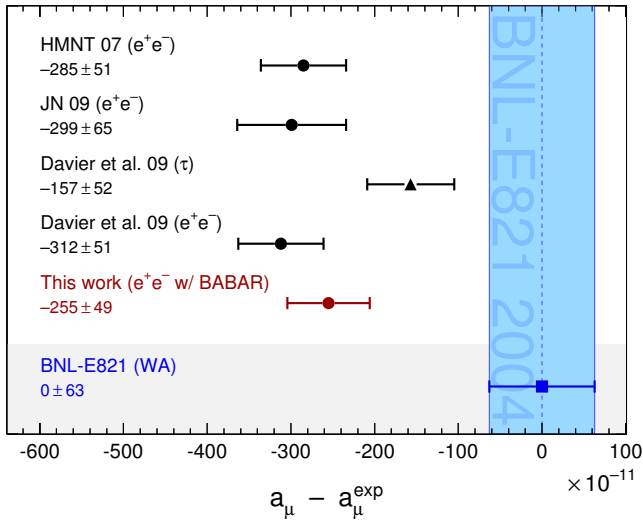}
\vspace{-0.4cm}
\caption{Compilation of recent results for $\amuSM$,
        subtracted by the central value of the experimental average~\cite{bnl}.
        The shaded vertical band indicates the experimental error. 
        The SM predictions are taken from: HMNT 07~\cite{hmnt}, JN 09~\cite{jeger},
        Davier \ea 09~\cite{eetaunew} (\Tau-based and \ee including KLOE), 
        and the \ee-based value from this work. }
\label{fig:amures}
\end{figure}

Also given in Table~\ref{tab:results} is the combined \Tau-based result from 
Ref.~\cite{eetaunew}. The difference between the \Tau and \ee-based evaluations of 
\amuhadLOpp now reads $6.8 \pm 3.5_{\tau+{\rm IB}} \pm 2.9_{ee}$, thus
reducing to $1.5\sigma$ compared to $2.4\sigma$ without BABAR~\cite{eetaunew}
(the BABAR-only result is in excellent agreement with the \Tau data).\footnote
{
   Combining the \ee and \Tau-based evaluations would lead to an error rescaling 
   by a factor of 1.5 to account for the inconsistency in the integrated data. 
   This would approximately cancel the expected precision gain from the combination. 
} 
A comparison between the combined \ee and \Tau two-pion cross sections relative to 
the \ee result is shown in Fig.~\ref{fig:eetau}. Significant local discrepancies arise 
in particular above the $\rho$ peak. 

We also reevaluate the $\ee\to\pp2\piz$ contribution to \amuhadLO. 
The CMD2 data used previously~\cite{cmd2pp2pi0} have been superseded by modified or 
more recent, but yet unpublished data~\cite{logashenko}, recovering agreement with the 
published SND cross sections~\cite{sndpp2pi0}. Since the new data are unavailable, we 
discard the obsolete CMD2 data from the $\pp2\piz$ average, finding 
$\amuhadLOppzz=17.6\pm0.4_{\rm stat}\pm1.7_{\rm syst}$ (compared to 
$17.0 \pm 0.4_{\rm stat} \pm 1.6_{\rm syst}$ when including the obsolete CMD2 data). 
The corresponding cross section measurements and HVPTools 
average are shown in Fig.~\ref{fig:4pi}.

Adding to the \ee-based \amuhadLOpp and \amuhadLOppzz results the remaining exclusive 
multi-hadron channels as well as perturbative QCD~\cite{md_tau06}, we find for the 
complete lowest order hadronic term
\beqns
   \amuhadLO[\ee] = 695.5 \pm 4.0_{\rm exp}\pm  0.7_{\rm QCD}~(4.1_{\rm tot})\,.
\eeqns
It is noticeable that the error from the \pp channel now equals the one from 
all other contributions to \amuhadLO.

Adding further the contributions from higher order hadronic loops, 
$-9.79 \pm 0.08_{\rm exp} \pm 0.03_{\rm rad}$~\cite{hmnt}, hadronic light-by-light 
scattering (LBLS), $10.5\pm 2.6$~\cite{prades09}, as well as QED,  
$11\,658\,471.809 \pm 0.015$~\cite{kinoshita} (see also \cite{pdgg-2rev} and 
references therein), and electroweak effects,
$15.4 \pm 0.1_{\rm had} \pm 0.2_{\rm Higgs}$~\cite{jackiw,czarnecki},
we obtain the SM prediction (still in $10^{-10}$ units)
\beqns
  \amuSM[\ee] &=& 11\,659\,183.4 \pm 4.1 \pm 2.6 \pm 0.2~(4.9_{\rm tot})\,,
\eeqns
where the errors have been split into lowest and higher order hadronic, and 
other contributions, respectively. The $\amuSM[\ee]$ value deviates from the 
experimental average, $\amuExp=11\,659\,208.9 \pm 5.4 \pm 3.3$~\cite{bnl,pdgg-2rev}, 
by $25.5 \pm 8.0$ ($3.2\sigma$).

A compilation of recent SM predictions for \amu compared with the experimental
result is given in Fig.~\ref{fig:amures}. The BABAR results are not yet contained
in evaluations preceding the present one. The result by HMNT~\cite{hmnt} contains 
older KLOE data~\cite{kloe04}, which have been superseded by more recent 
results~\cite{kloe08}, leading to a slightly larger value for \amuhadLO.

\section{Conclusions}

We have reevaluated the lowest order hadronic contribution to the muon 
magnetic anomaly in the dominant \pp channel, using new precision data published 
by the BABAR Collaboration. After combination with the other \ee  data a 
$1.5\sigma$ difference with the \Tau data remains for the dominant \pp contribution. 
For the full \ee-based Standard Model prediction, including also a reevaluated
$\pp2\piz$ contribution, we find a deviation of $3.2\sigma$ from experiment
(reduced from $3.7\sigma$ without BABAR). The deviation reduces to $2.9\sigma$ 
when excluding KLOE data, and further decreases to $2.4\sigma$ when using only 
the BABAR data in the \pp channel. As a reminder, the \Tau-based result deviates
by $1.9\sigma$ from the Standard Model.

The present situation for the evaluation of \amuhadLOpp is improved compared to 
that of recent years, as more input data from quite different experimental facilities 
and conditions have become available: \ee energy scan, \ee ISR from low and high energies, 
\Tau decays. Our attitude has been to combine all the data and include in the 
uncertainty the effects from differences in the spectra. At the moment the ideal 
accuracy cannot be reached as a consequence of the existing discrepancies due to 
uncorrected or unaccounted systematic effects in the data. A critical look must
be given to the different analyses in order to identify their weak points and to 
improve on them or to assign larger systematic errors.

It is thereby not sufficient to concentrate on improving the \pp channel
alone. Problems also persist in the $\pp2\piz$ mode, where the \Tau
and \ee-based evaluations differ by $(3.8\pm2.2)\cdot10^{-10}$, but also the 
\ee data among themselves exhibit discrepancies. Fortunately, new precision 
data from BABAR should soon help to clarify the situation in that channel. 

\begin{details}
This work has been supported in part by the National Natural Science 
Foundation of China (10825524) and the Talent Team Program of CAS (KJCX2-YW-N45).
\end{details}

\end{document}

%% file: defs.tex
\mathchardef\Upsilon="7107
\def\Y#1S{\ensuremath{\Upsilon{(#1S)}}\xspace}

\newcommand{\Kbar    }{\kern 0.2em\overline{\kern -0.2em K}{}\xspace}

\newcommand{\Kz      }{\ensuremath{K^0}\xspace}
\newcommand{\Kzb     }{\ensuremath{\Kbar^0}\xspace}
\newcommand{\KzKzb   }{\ensuremath{\Kz \kern -0.16em \Kzb}\xspace}
\newcommand{\Kp      }{\ensuremath{K^+}\xspace}
\newcommand{\Km      }{\ensuremath{K^-}\xspace}

\newcommand{\KpKm    }{\ensuremath{\Kp \kern -0.16em \Km}\xspace}

\newcommand{\Tau}{\ensuremath{\tau}\xspace}
\newcommand{\nut}{\ensuremath{\nu_\tau}\xspace}

\newcommand{\BR}{\ensuremath{{\cal B}}\xspace}

\newcommand{\pip}{\ensuremath{\pi^+}\xspace}
\newcommand{\pim}{\ensuremath{\pi^-}\xspace}
\newcommand{\piz}{\ensuremath{\pi^0}\xspace}

\newcommand{\ee}{\ensuremath{e^+e^-}\xspace}
\newcommand{\mm}{\ensuremath{\mu^+\mu^-}\xspace}
\newcommand{\mmg}{\ensuremath{\mu^+\mu^-(\gamma)}\xspace}

\newcommand{\pp}{\ensuremath{\pi^+\pi^-}\xspace}

\newcommand{\ppg}{\ensuremath{\pi^+\pi^-(\gamma)}\xspace}

\newcommand{\tev}{\ensuremath{\mathrm{\,Te\kern -0.1em V}}\xspace}
\newcommand{\gev}{\ensuremath{\mathrm{\,Ge\kern -0.1em V}}\xspace}
\newcommand{\mev}{\ensuremath{\mathrm{\,Me\kern -0.1em V}}\xspace}
\newcommand{\kev}{\ensuremath{\mathrm{\,ke\kern -0.1em V}}\xspace}
\newcommand{\ev}{\ensuremath{\mathrm{\,e\kern -0.1em V}}\xspace}
\newcommand{\gevc}{\ensuremath{{\mathrm{\,Ge\kern -0.1em V\!/}c}}\xspace}
\newcommand{\mevc}{\ensuremath{{\mathrm{\,Me\kern -0.1em V\!/}c}}\xspace}
\newcommand{\gevcc}{\ensuremath{{\mathrm{\,Ge\kern -0.1em V\!/}c^2}}\xspace}
\newcommand{\mevcc}{\ensuremath{{\mathrm{\,Me\kern -0.1em V\!/}c^2}}\xspace}

\newcommand{\bei}{\begin{itemize}}
\newcommand{\eei}{\end{itemize}}
\newcommand{\ben}{\begin{enumerate}}
\newcommand{\een}{\end{enumerate}}
\newcommand{\beq}{\begin{equation}}
\newcommand{\eeq}{\end{equation}}
\newcommand{\beqn}{\begin{eqnarray}}
\newcommand{\eeqn}{\end{eqnarray}}
\newcommand{\beqns}{\begin{eqnarray*}}
\newcommand{\eeqns}{\end{eqnarray*}}

\renewcommand{\arraystretch}{1.25}
\newcommand{\amu}{\ensuremath{a_\mu}\xspace}

\newcommand{\amuhadLO}{\ensuremath{\amu^{\rm had,LO}}\xspace}
\newcommand{\amuhadLOpp}{\ensuremath{\amu^{\rm had,LO}[\pi\pi]}\xspace}
\newcommand{\amuhadLOppzz}{\ensuremath{\amu^{\rm had,LO}[\pi\pi2\piz]}\xspace}
\newcommand{\amuhadHO}{\ensuremath{\amu^{\rm had,HO}}\xspace}
\newcommand{\amuSM}{\ensuremath{\amu^{\rm SM}}\xspace}
\newcommand{\amuExp}{\ensuremath{\amu^{\rm exp}}\xspace}
\newcommand\ie{{\it i.e.}\xspace}

\newcommand\cf{{\em cf.}\xspace}
\newcommand{\ea}{{\em et al.}\xspace}
\def\@citex[#1]#2{\if@filesw\immediate\write\@auxout{\string\citation{#2}}\fi
  \@tempcnta\z@\@tempcntb\m@ne\def\@citea{}\@cite{\@for\@citeb:=#2\do
    {\@ifundefined
       {b@\@citeb}{\@citeo\@tempcntb\m@ne\@citea
        \def\@citea{,\penalty\@m\ }{\bf ?}\@warning
       {Citation `\@citeb' on page \thepage \space undefined}}%
    {\setbox\z@\hbox{\global\@tempcntc0\csname b@\@citeb\endcsname\relax}%
     \ifnum\@tempcntc=\z@ \@citeo\@tempcntb\m@ne
       \@citea\def\@citea{,\penalty\@m}
       \hbox{\csname b@\@citeb\endcsname}%
     \else
      \advance\@tempcntb\@ne
      \ifnum\@tempcntb=\@tempcntc
      \else\advance\@tempcntb\m@ne\@citeo
      \@tempcnta\@tempcntc\@tempcntb\@tempcntc\fi\fi}}\@citeo}{#1}}

\def\@citeo{\ifnum\@tempcnta>\@tempcntb\else\@citea
  \def\@citea{,\penalty\@m}%
  \ifnum\@tempcnta=\@tempcntb\the\@tempcnta\else
   {\advance\@tempcnta\@ne\ifnum\@tempcnta=\@tempcntb \else
\def\@citea{--}\fi
    \advance\@tempcnta\m@ne\the\@tempcnta\@citea\the\@tempcntb}\fi\fi}
\newenvironment{myquote}
               {\list{}{\leftmargin0cm\indent}%
                \item\relax}
               {\endlist}
\newcommand\allFontSize{\footnotesize}
\newcommand\detailsSize{\allFontSize}
\newenvironment{details}%
{\begin{myquote}\detailsSize}{\end{myquote}}


%% file: g209.bbl
\begin{thebibliography}{99}
\bibitem{hvptools}   The HVPTools source code and database 
                     can be made publicly available. Please contact the authors.
\end{thebibliography}
